\documentclass[a4paper,draft]{book}
\usepackage{epsf}
\usepackage{nano2cmr}
\usepackage{textcomp}

\begin{document}

\pnum{}
\ttitle{Ge/Si(001) heterostructures with quantum dots: formation, defects, 
photo-electromotive force and terahertz conductivity}
\tauthor{{\em V.~A.~Yuryev}, {L.~V.~Arapkina}, M.~S.~Storozhevykh,  V.~A.~Chapnin, K.~V.~Chizh, O.~V.~Uvarov, V.~P.~Kalinushkin, E.~S.~Zhukova, A.~S.~Prokhorov, I.~E.~Spektor and B.~P.~Gorshunov}

\ptitle{Ge/Si(001) heterostructures with quantum dots: formation,\\ defects, 
photo-electromotive force and terahertz conductivity}
\pauthor{{\em V.~A.~Yuryev}$^{1,2}$, {L.~V.~Arapkina}$^{1}$, M.~S.~Storozhevykh$^{1}$,  V.~A.~Chapnin$^{1}$, K.~V.~Chizh$^{1}$, O.~V.~Uvarov$^{1}$, V.~P.~Kalinushkin$^{1,2}$, E.~S.~Zhukova$^{1,3}$, A.~S.~Prokhorov$^{1,3}$, I.~E.~Spektor$^{1}$ and B.~P.~Gorshunov$^{1,3}$}

\affil{$^{1}$~Prokhorov General Physics Institute of the Russian Academy of Sciences, 38 Vavilov Street, Moscow, 119991, Russia
\\ $^{2}$~Technopark of GPI RAS, 38 Vavilov Street, Moscow, 119991, Russia
\\ $^{3}$~Moscow Institute of Physics and Technology, Institutsky Per. 9, Dolgoprudny, Moscow Region, 141700, Russia}

\begin{abstract}
{Issues of Ge hut cluster nucleation and growth at low temperatures on the Ge/Si(001) wetting layer are discussed on the basis of explorations performed by high resolution STM and {\it in-situ} RHEED. Data of HRTEM investigations of  Ge/Si heterostructures  are presented with the focus on low-temperature formation of perfect multilayer films. 
Exploration of the photovoltaic effect in Si {\it p--i--n}-struc\-tu\-res with Ge quantum dots allowed us to propose a new approach to designing of infrared detectors. 
First data on THz dynamical conductivity of Ge/Si(001) heterostructures in the temperature interval from 5 to 300\,K  and magnetic fields up to 6\,T are reported.}
\end{abstract}

\begindc 

\index{Yuryev V. A.}
\index{Arapkina L. V.}   
\index{Storozhevykh M. S.}
\index{Chapnin V. A.}
\index{Chizh K. V.}
\index{Uvarov O. V.}
\index{Kalinushkin V. P.}
\index{Zhukova E. S.}
\index{Prokhorov A. S.}
\index{Spektor I. E.}
\index{Gorshunov B. P.}

\section*{Introduction}

Results of our researches of Ge/Si(001) structures aimed at development of CMOS compatible IR detectors have been published in a number of publications [1--4].
This article reports our recent results concerning formation of Ge quantum dot arrays, growth of multilayer structures, their photoelectrical properties and dielectric characteristics at THz frequencies.

\section {Formation and defects}

We  demonstrate that huts form via parallel nucleation of two characteristic embryos different only in symmetry and composed by epitaxially oriented Ge dimer pairs and  chains of four dimers on tops of the wetting layer (WL) $M\times N$ patches: an individual embryo for each species of huts---pyramids or wedges. These nuclei always  arise on sufficiently large WL patches: there must be enough space for a nucleus on a single patch; a nucleus cannot form on more than one patch. This fact may be explained in assumption of presence of the Ehrlich-Schwoebel barriers on sides of  WL patches which prevent the spread of dimer chains composing nuclei from one patch to another; nuclei form on bottoms of potential wells of the spatial potential relief associated with  the WL patches. The total patch thickness (from Si/Ge interface to the patch top) rather than the mean thickness of WL controls the hut nucleation process. A growing hut likely reduces and finally eliminates the potential barrier; then it occupies adjacent WL patches.

We suppose also that WL patch top reconstruction determines the type of a hut which can form on a patch. If this is the case, the $c(4\times 2)$ reconstruction of the patch top enables  the nucleation of a pyramid, whereas the $p(2\times 2)$ reconstruction allows a wedge to arise on the patch. Domination of one of these reconstruction types may result in domination of a certain species of huts.

Dynamics of the RHEED patterns in the process of Ge hut array formation is investigated at low and high temperatures of Ge deposition. 
At high temperatures, as $h_{\rm Ge}$ increases, diffraction patterns evolve as $(2\times 1)\rightarrow (1\times 1)\rightarrow (2\times 1)$ with very weak {\textonehalf}-reflexes. Brightness of the {\textonehalf}-reflexes increases (the $(2\times 1)$ structure becomes pronounced) and the 3D-reflexes arise only during sample cooling. At low temperatures, the structure changes as $(2\times 1)\rightarrow (1\times 1)\rightarrow (2\times 1)\rightarrow (2\times 1)+3$D-reflexes.
Different dynamics of RHEED patterns during Ge deposition  in different growth modes  reflects the difference in mobility  and  `condensation' fluxes from Ge 2D gas of  Ge adatoms on the surface, which in turn control the nucleation rates and densities of Ge clusters. 
 High Ge mobility and low cluster nucleation rate in comparison with fluxes to competitive  sinks of adatoms determine the observed difference in the surface structure formation at high temperatures as compared with that at low temperatures.

\begin{figure}[h]
\leavevmode
\centering{\epsfbox{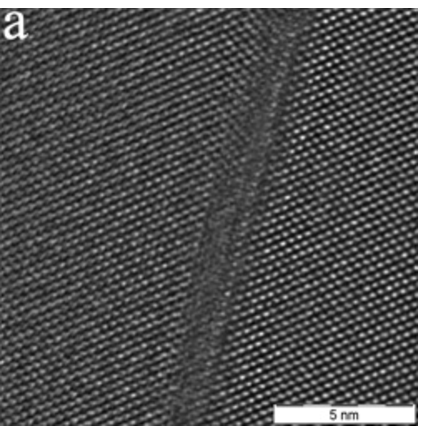}}
\centering{\epsfbox{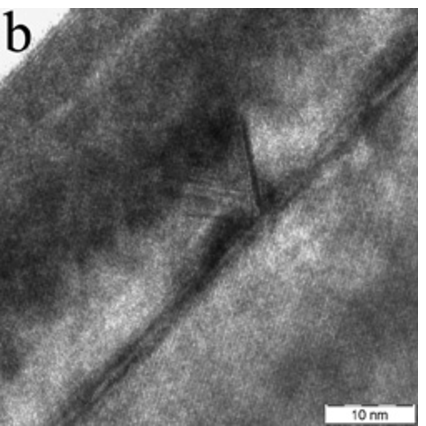}}
\caption{
HRTEM micrographs of Ge quantum dots in Si; $T_{gr}=360${\textcelsius} for Ge and 530{\textcelsius} for Si,  $h_{\rm Ge}$  is (a) 6\,\r{A}, (b) 10\,\r{A}.}
\end{figure}

\begin{figure}[h]
\leavevmode
\centering{\epsfbox{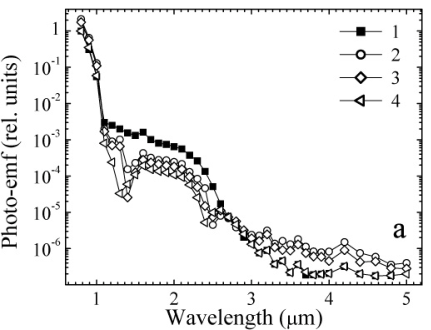}}~
\centering{\epsfbox{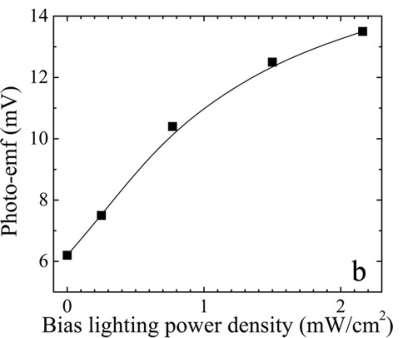}}
\caption{
(a) Photo-emf spectra of the {\it p--i--n-}structure;
(1)~without bias lighting; 
(2)--(4) under wide-band IR bias lighting  (tungsten bulb, Si filter, $\lambda > 1.1\,\mu$m):
(2)~$W=0.63$\,mW/cm$^2$;
(3)~$W=5.3$\,mW/cm$^2$;
(4)~$W=17.5$\,mW/cm$^2$.
(b) The dependence of the {\it p--i--n-}structure photo-emf, generated by narrow-band illumination in Si fundamental absorption range,  on the  power density of  bias lighting; $T=80$\,K.}
\end{figure}

Data of HRTEM studies evidence that  extended defects do not arise at low Ge coverages ($h_{\rm Ge}$) on the buried Ge clusters (Fig.~1a)
and perfect epitaxial heterostructures with quantum dots form under these conditions that enables the formation of  defectless  multilayer structures suitable for device applications. Stacking faults (SF) have been found to arise on Ge clusters at $h_{\rm Ge}$ as large as 10\,\r{A} (Fig.~1b).
Multiple SFs often damage multilayer structures at these $h_{\rm Ge}$.



\section{Photo-electromotive force}

Heteroepitaxial Si {\it p--i--n}-diodes with multilayer stacks of Ge/Si(001) quantum dot dense arrays built in intrinsic domains have been investigated and found to exhibit the pho\-to-emf in a wide spectral range from 0.8 to 5 $\mu$m. The diodes comprised 
$n$-Si(100) substrate  ($0.1\,\Omega$\,cm), Si buffer (1690\,nm), 4 periods of [Ge (1\,nm), Si barrier (30\,nm)], [Ge (1\,nm), Si (50\,nm)], $p$-Si:B cap (212\,nm, $p\sim 10^{19}$\,cm$^{-3}$)
An effect of wide-band irradiation by IR light on the photo-emf spectra has been observed. Pho\-to-emf in different spectral ranges has been found to be differently affected by the wide-band lighting (Fig.~2a).
A significant increase in photo-emf is observed in the fundamental absorption range under the wide-band IR irradiation (Fig.~2b).
The above phenomena are explained in terms of positive and neutral charge states of the quantum dot layers and the Coulomb potential of the quantum dot ensemble which changes under the wide-band IR radiation.   

A new design of photovoltaic quantum dot infrared photodetectors is proposed [5]
which enables detection of variations of photo-emf produced by the narrow-band radiation in the Si fundamental absorption range under the effect of the wide-band IR radiation resulted from changes in the Coulomb potential of the quantum dot ensemble affecting the efficiency of the photovoltaic conversion. The quantum dot array resembles a grid of a triode in these detectors which is controlled by the detected IR light. The reference narrow-band radiation generates a potential between anode and cathode of this optically driven quantum dot triode. Such detectors can be realized on the basis of any appropriate semiconductor structures with potential barriers, e.g., $p$--$n$-junctions or Schottky barriers, and built-in arrays of nanostructures.


\section{Terahertz conductivity}

\begin{figure}[t]
\leavevmode
\centering{\epsfbox{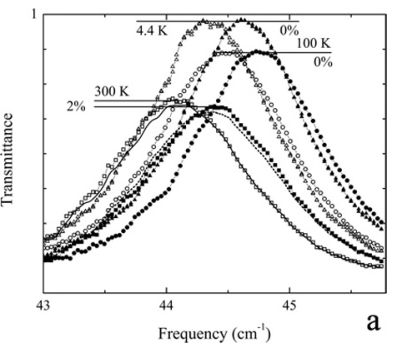}}~
\centering{\epsfbox{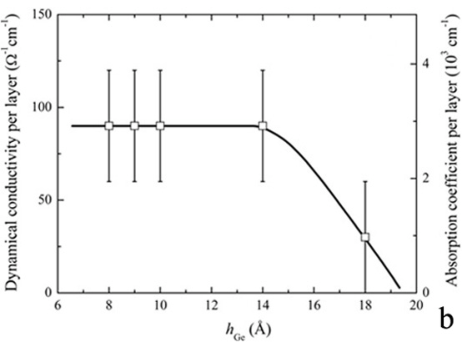}}
\caption{
(a) Transmittance spectra of a 5-layer Ge/Si(001) heterostructure ($T_{\rm gr}=360^{\circ}$C, $h_{\rm Ge}=8${\,\AA}) recorded at different temperatures; dark and open symbols relate to substrate regions with and without the heterostructure; solid and dotted lines show spectra obtained  at 300\,K under magnetic field of 6\,T directed normal to the surface, without and with the heterostructure, respectively. (b) Terahertz conductivity and absorption coefficient of terahertz radiation in multilayer Ge/Si(001) heterostructures vs Ge coverage in each layer; $\nu = 1$\,THz, $T=300$\,K.}
\end{figure}

By using a coherent source spectrometer [6],
first measurements of THz dynamical conductivity (or absorptivity) spectra of Ge/Si(001) heterostructures were performed at frequencies from 0.3 to 1.2\,THz in the temperature interval from 300 to 5\,K and also in magnetic fields up to 6\,T. The effective dynamical conductivity of the heterostructures was determined by comparing the terahertz transmission coefficient spectra of the sample with the heterostructure on Si substrate with that of the same sample but with the heterostructure etched away.  The results are exemplified by the transmission coefficients spectra shown in Fig.~3a.
Although rather small, the difference in the amplitudes of the interference maxima (due to multiple reflections of the monochromatic radiation between the faces of the sample) is clearly seen and reliably detected. Processing this kind of spectra allowed us to extract the dynamical conductivity of heterostructures separately and  trace its dependence on the thickness of the germanium coverage layer; the latter determines whether this layer organizes itself into an array of quantum dots or stays as a uniform layer. As a results, the THz dynamical conductivity of Ge quantum dots (Fig.~3b)
has been discovered to be significantly higher than that of the structure with the same amount of bulk germanium (not organized in an array of quantum dots). The excess conductivity is not observed in the structures with the Ge coverage less than 8\,\AA. When a Ge/Si(001) sample is cooled down the conductivity decreases (Fig.~4).
It is seen from Fig.~3a
that the samples display positive magnetoresistivity.

\begin{figure}[t]
\leavevmode
\centering{\epsfbox{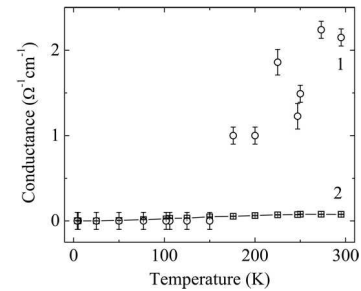}}
\caption{
Temperature dependences of THz conductance of the  Ge/Si(001) heterostructure (1) and the Si substrate (2).}
\end{figure}


\acks  This research has been supported by the Ministry of Education and Science of RF through the contracts No.~14.740.11. 0069 and No. 16.513.11.3046 and RFBR through the grant No. 11-02-12023-ofi-m. Center of Collective Use of Scientific Equipment of GPI RAS is appreciated for instrumentation and Prof. M. Dressel for providing equipment for magnetic measurements.


\begin{thebibliography}{8}
\itemsep-2pt

\bibitem{Yur1-PRB2010} 
L. V. Arapkina \etal, {\em Phys. Rev. B} {\bf 82}, 045315 (2010).



\bibitem{Yur1-JAP2011} 
L.~V.~Arapkina \etal, {\em J. Appl. Phys.} {\bf 109}, 104319 (2011).

\bibitem{Yur1-NRL2011-EMRS} 
L.~V.~Arapkina \etal, {\em Nanoscale Res. Lett.} {\bf 6}, 345 (2011).

\bibitem{Yur1-NRL2011-VCIAN} 
V.~A.~Yuryev \etal, {\em Nanoscale Res. Lett.} {\bf 6}, 522 (2011).

\bibitem{Yur1-patent-Ge} 
V.~A.~Yuryev \etal, Patent pending (2012).

\bibitem{Yur1-Gorshunov} 
B. Gorshunov \etal, {\em Int. J. Infrared Millimeter Waves} {\bf 26}, 1217 (2005).


\end{thebibliography}
\end{document}